\documentclass[english]{article}
\usepackage{amsfonts}
\usepackage{amsmath}
\usepackage{amssymb}
\usepackage{amsthm}
\usepackage{verbatim}
\usepackage{float}
\usepackage{subfig}
\usepackage{multirow}
\usepackage[colorlinks,citecolor=blue,urlcolor=blue,linkcolor=red]{hyperref}
\usepackage{fullpage}
\usepackage[affil-it]{authblk}
\usepackage{siunitx}
\usepackage{textcomp}
 \usepackage{tabularx}
\usepackage{booktabs}
\usepackage{caption}
\usepackage{graphicx}

\renewcommand{\phi}{\varphi}
\renewcommand{\epsilon}{\varepsilon}

\begin{document}
\title{OCT-GAN: Single Step Shadow and Noise Removal from Optical Coherence Tomography Images of the Human Optic Nerve Head}

\author[1, 2]{Haris Cheong}
\author[1]{Sripad Krishna Devalla}
\author[1, 2]{Thanadet Chuangsuwanich}
\author[3]{Tin A. Tun}
\author[4]{Xiaofei Wang}
\author[1,3]{Tin Aung}
\author[3, 5, 6, 7, 8, 9, 10]{Leopold Schmetterer}
\author[2]{Martin L. Buist}
\author[1,11]{Craig Boote}
\author[12]{Alexandre H. Thi\'{e}ry}
\author[1,3 $\star$]{Micha\"el J. A. Girard}

\affil[1]{Ophthalmic Engineering \& Innovation Laboratory (OEIL), Singapore Eye Research Institute, Singapore National Eye Centre, Singapore}
\affil[2]{Department of Biomedical Engineering, Faculty of Engineering, National University of Singapore, Singapore}
\affil[3]{Singapore Eye Research Institute, Singapore National Eye Centre, Singapore}
\affil[4]{Beijing Advanced Innovation Center for Biomedical Engineering, School of Biological Science and Medical Engineering, Beihang University, Beijing, China}
\affil[5]{Ophthalmology Department. Duke-NUS Medical School, Singapore}
\affil[6]{School of Clinical and Biomedical Engineering. Nanyang Technological University, Singapore}
\affil[7]{SERI-NTU Advanced Ocular Engineering (STANCE), Singapore, Singapore}
\affil[8]{Center for Medical Physics and Biomedical Engineering, Medical University Vienna, Austria}
\affil[9]{Department of Clinical Pharmacology, Medical University of Vienna, Austria}
\affil[10]{Institute of Molecular and Clinical Ophthalmology, Basel, Switzerland}
\affil[11]{Structural Biophysics Group, School of Optometry \& Vision Sciences, Cardiff University, UK}
\affil[12]{Department of Statistics and Applied Probability, National University of Singapore, Singapore}
\affil[$\star$]{mgirard@ophthalmic.engineering}

\bigskip

\maketitle


%
%
\begin{abstract}
\noindent
Abstract: Speckle noise and retinal shadows within OCT B-scans occlude important edges, fine textures and deep tissues, preventing accurate and robust diagnosis by algorithms and clinicians. We developed a single process that successfully removed both noise and retinal shadows from unseen single-frame B-scans within 10.4ms. Mean average gradient magnitude (AGM) for the proposed algorithm was 57.2\% higher than current state-of-the-art, while mean peak signal to noise ratio (PSNR), contrast to noise ratio (CNR), and structural similarity index metric (SSIM) increased by 11.1\%, 154\% and 187\% respectively compared to single-frame B-scans. Mean intralayer contrast (ILC) improvement for the retinal nerve fiber layer (RNFL), photoreceptor layer (PR) and retinal pigment epithelium (RPE) layers decreased from 0.362 ± 0.133 to 0.142 ± 0.102, 0.449 ± 0.116 to 0.0904 ± 0.0769, 0.381 ± 0.100 to 0.0590 ± 0.0451 respectively. The proposed algorithm reduces the necessity for long image acquisition times, minimizes expensive hardware requirements and reduces motion artifacts in OCT images.

\end{abstract}

\section{INTRODUCTION}
\label{sec:intro}
Optical coherence tomography (OCT) is a well-established, noninvasive clinical imaging tool for in vivo viewing of cross-sectional images of optical nerve head (ONH) tissues with micrometer resolution \cite{RN54}. Although there have been vast improvements in imaging resolution, speed, and depth of OCT imaging, some limitations exist. Since OCT uses coherent illumination, speckle noise is a major source of noise that degrades the image quality of OCT B-scans \cite{RN13}.\\

Speckle noise is a multiplicative noise inherent in coherence imaging and is caused by multiple forward and backward scattering of light waves. It frequently reduces contrast and the grainy speckle noise pattern has been found to limit both the axial and lateral effective image resolution \cite{RN48}. Subtle but important morphological details, such as individual tissue layers \cite{RN33, RN34, RN35} are prevented from being identified and observed \cite{RN39}, making speckle noise detrimental to clinical diagnosis \cite{RN1}. \\

The most common speckle removal approach adopted in commercial OCT machines is B-scan averaging \cite{RN58}. Although high quality images can be produced using this technique, the longer scan durations required for this technique cause other artifacts such as registration errors \cite{RN17}, motion artifacts \cite{RN18} to appear on processed images due to eye or head motion \cite{RN50}. The inability of elderly or young patients to remain fixated for long periods of time further render this technique difficult to obtain 3D scans of the ONH \cite{RN60} of relatively good quality. \\

Furthermore, multi-frame averaging does not prevent OCT signals obtained from locations beneath retinal blood vessels from being significantly diminished due to the scattering at the blood flowing through retinal blood vessels. This phenomenon produces artifacts in OCT images known as retinal shadows. These artifacts appear perpendicular to retinal layers, interrupting tissue layer continuity and causing errors in segmentation \cite{RN23}.  This in turn leads to inaccurate extraction of important structural metrics such as thickness of the retinal nerve fiber layer (RNFL), which is important in glaucoma monitoring \cite{RN24}. Retinal shadows also reduce visibility of deep structures such as the anterior and posterior boundaries of the lamina cribrosa (LC), as weak, reflected signals from these structures are further attenuated by the lower incident light intensity within retinal shadows \cite{RN2}.
Recently, deep learning techniques have shown promise in reducing speckle noise. Mao et al. used a deep, fully convolutional encoding-decoding framework to suppress noise and perform super resolution analysis of input images \cite{RN20}. Later in 2018, Ma et al. proposed an edge-sensitive generative adversarial network (GAN) to remove speckle noise from OCT images produced by commercial scanners \cite{RN21}. Devalla et al. leveraged deep neural networks (DNNs), residual learning, and dilated convolutions to extract multi-scale features and contextual information to recover information lost due to speckle noise in OCT images of the ONH \cite{RN32}. Many other works attempted to remove speckle noise with varying success, with a common recognition of the major quality degrading factor that speckle noise inflict on OCT images \cite{RN40, RN41, RN42}.\\

Some have attempted to remove retinal shadows as well. In 2011, Girard et al. developed two OCT modelling approaches to be used in conjunction, one to compensate for light attenuation and the other to enhance contrast in OCT images \cite{RN2}. Later, in 2018, Vupparaboina et al. illustrated an improvement in choroid representation after shadow compensation \cite{RN26}. Our more recent work \cite{RN27} used a weighted custom loss function that removed shadows from multi-frame averaged images and illuminated faint features within retinal shadows. However, the above-mentioned algorithms require high quality images free from speckle noise and motion artifacts to function well, preventing users in possession of single-frame images and low-cost hardware from availing themselves to this technology. \\

The presence of speckle noise, motion artifacts, and retinal shadows often interact and overlap, complicating processes that attempt to alleviate and remove these quality degrading phenomena \cite{RN33, RN63}. Such attempts are often tedious and prone to errors, because multiple separate processes need to work together to remove each artifact individually, with the ordering of artifact removal causing issues for the other processes. In this study, we aimed to develop an algorithm to remove both speckle noise and retinal shadows within a single step. By doing so, we will be able to reduce the cost of OCT devices by using simpler OCT imaging hardware enhanced by software.

\section{METHODS}
\label{sec:methods}
\subsection{Patient Recruitment}
24 healthy subjects (average age: 28 years) were recruited at the Singapore National Eye Centre (SNEC). All subjects gave written informed consent. This study adhered to the tenets of the Declaration of Helsinki and was approved by the institutional review board of the hospital. The inclusion criteria for healthy subjects were an intraocular pressure (IOP) of less than 21mmHg and healthy optic nerves with a vertical cup-to-disc ratio of $\leq$ 0.5.

\subsection{OCT Imaging}
Recruited subjects were seated and imaged in dark room conditions by a single operator (TAT). A standard spectral domain OCT system (Spectralis; Heidelberg Engineering, Heidelberg, Germany) was used to image both eyes of each subject. Each volume contained 97 horizontal B-scans (32-µm distance between B-scans; 384 A-scans per B-scan) from a rectangular area 15° × 10° centered on the ONH.  We obtained a total of 2628 single-frame B-scans (noisy, without signal averaging) and 2628 multi-frame B-scans (clean, averaged over 75 frames). Enhanced depth imaging (EDI) \cite{RN69} and eye tracking \cite{RN70, RN72} modalities were used during the acquisition.

\subsection{Overall description}
Our algorithm was a single step approach to removing both speckle noise and retinal blood vessel shadows simultaneously. It had two actively-trained networks competing with one another. The first network was referred to as the shadow detector network, and it predicted which pixels would be considered as shadowed pixels. The second network was referred to as the image processor, and it aimed to remove shadows and speckle noise simultaneously from single-frame OCT images such that the first network (shadow detection network) could no longer identify shadowed pixels. Briefly, we trained the shadow detection network once on multi-frame images with added Gaussian noise with their corresponding manually segmented shadow mask as the ground truth. We added Gaussian noise instead of speckle noise as training deep learning models to denoise B-scans with Gaussian noise provided empirically better results. It is unfortunately, difficult to ascertain the reason for this phenomenon.\\

\begin{figure}[H] 
    \centering
    \includegraphics[width=\textwidth,height=\textheight,keepaspectratio]{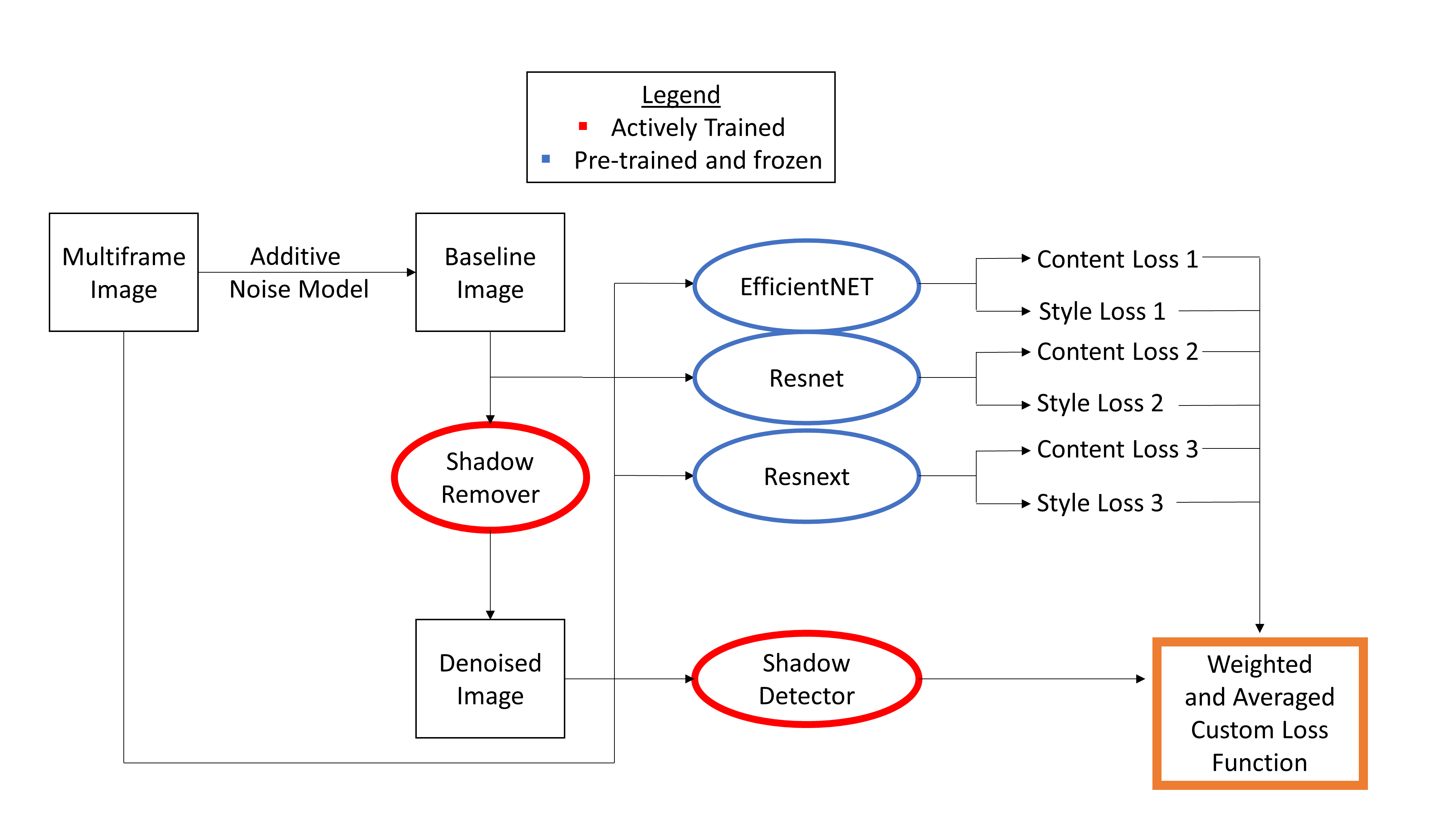}
    \caption{Overview of the proposed deep learning framework.}
    \label{fig:1}
\end{figure}

First, binary segmentation masks (size 496 × 384) were manually created for all 2328 training B-scans using ImageJ [29] by one observer (HC) where shadowed pixels were labelled as 1 and shadow-free pixels were labelled as 0. Next, we attempted to model single-frame images by creating “noisy” images. This was done by adding Gaussian noise to multi-frame images. Seven feature representations of each noisy image and its multi-frame counterpart were extracted using three pre-trained perceptual networks in order to train the image processor network to output multi-frame quality images from input noisy images. Finally, we trained the image processor network by passing the multi-frame image (with artificial Gaussian noise) as input and using the predicted binary masks as part of the loss function. More details about the overall algorithm can be found below (Fig. \ref{fig:1}).

\subsection{Shadow Detection Network and Image Processor Network Architecture}
\begin{figure}[H] 
    \centering
    \includegraphics[width=\textwidth,height=\textheight,keepaspectratio]{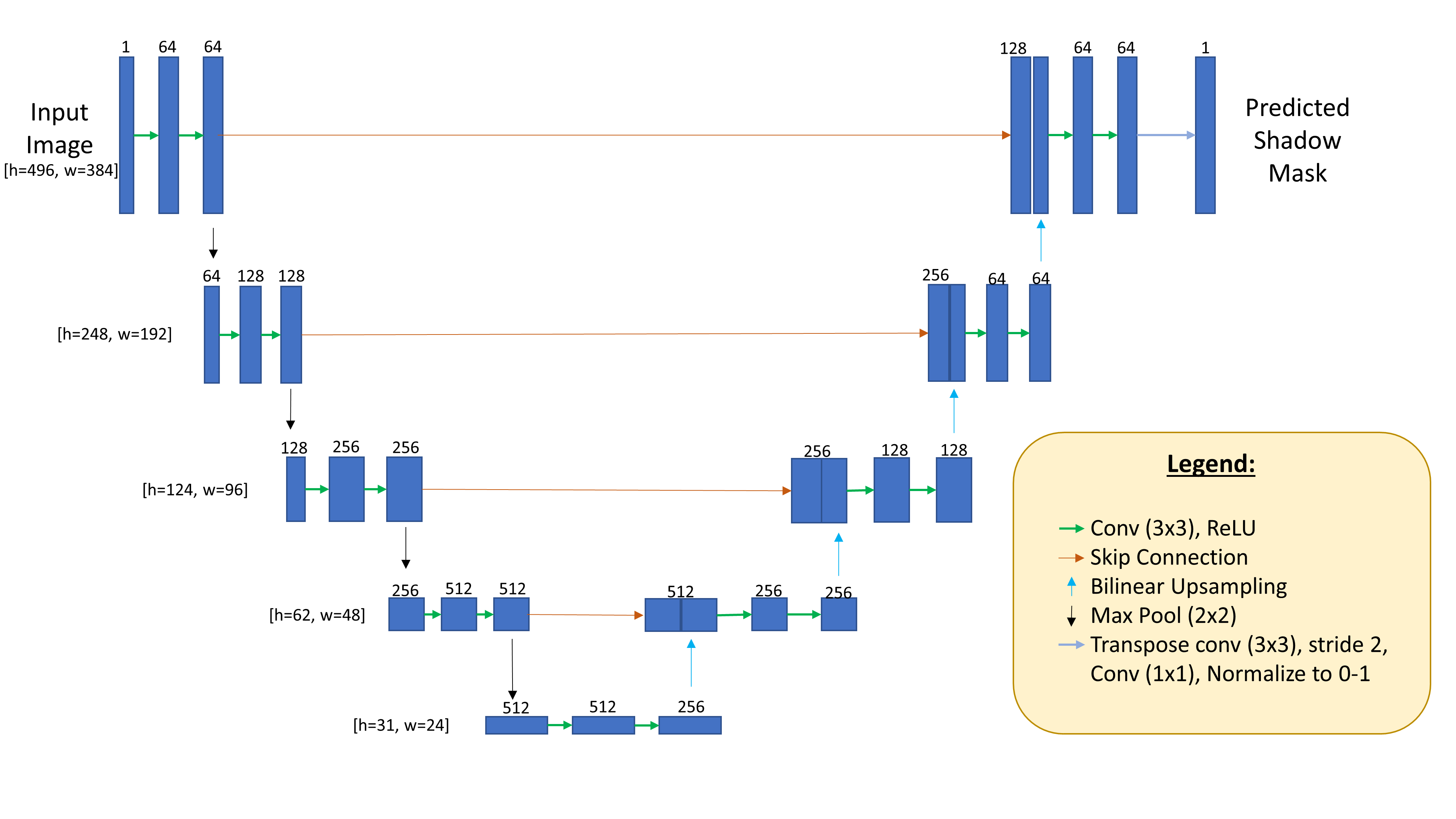}
    \caption{UNet Architecture used in the shadow detector network.}
    \label{fig:2}
\end{figure}

A neural network inspired by the UNet architecture \cite{RN64} (Fig. \ref{fig:2}) was trained with a simple binary cross entropy loss \cite{RN65}, using the noisy images (multi-frame + Gaussian noise) as inputs and the manually segmented masks as ground truths. This network had a sigmoid layer as its final activation, making it a per-pixel binary classifier. The shadow detector network first performed two convolutions with kernel size 3 and stride 1, followed by a ReLU activation \cite{RN66} after each convolution. Then, images were downsampled with a 2×2 kernel, halving the height and width of the feature maps. This occurred four times, with the number of feature maps at each smaller size increasing from 1 to 64, 128, 256, and 512, respectively. The shadow detection network was comprised of two towers. A downsampling tower halved the dimensions of the input image (size 512 × 512) via maxpooling to capture contextual information such as the spatial arrangement of tissues, and an upsampling tower sequentially restored it back to its original resolution to capture the local information such as tissue texture \cite{RN32}. Output images were then linearly scaled to values between 0 and 1 by subtraction of its minimum value and division by its maximum value.

\subsection{Image Augmentation}
To ensure that our algorithm was robust and functioned on single-frame images with varying levels of noise and retinal shadows, we implemented online image rotation (-45° to 45°), XY translation (-50\% to 50\% of image size), image scaling (-50\% to +50\% of image size) and random horizontal flip during our training.

\subsection{Speckle Noise Modelling}
We needed to add noise to multi-frame images to simulate the speckle noise found in single-frame images. The goal was to train the image processor to remove this artificial noise and in turn enable the image processor network to remove genuine speckle noise found in single-frame OCT images. We found through experiments that speckle noise was able to be modelled as Gaussian noise ($\mu$ = 0, $\sigma$ = 1) multiplied with a uniform distribution (range 0.02 to 0.5). In addition, including a large range for the Gaussian model helped the algorithm to perform robustly on single-frame images, which had varying levels of noise. These numbers were experimentally obtained by qualitative assessment of test images generated from single-frame images. A new noise sample was created for every multi-frame image during training to encourage robust training of the image processor network.

\subsection{Feature Extraction}
As using mean squared error (MSE) directly on processed images as a loss function was found to produce blurring effects on processed images, we instead applied MSE onto extracted feature representations of noisy images and their corresponding multi-frame B-scans. To extract comprehensive feature representations of each image, we required capable, pre-trained networks for feature extraction. Our framework consisted of three pre-trained and frozen feature extraction networks (Fig. \ref{fig:1}). These frozen networks were used to extract features from input images and will henceforth be referred to as perceptual networks. We used the three classification networks trained on ImageNet as our perceptual networks, namely, EfficientNet-B4 \cite{RN43}, WideResnet101\_2 \cite{RN44}, and Resnext101\_32x8d \cite{RN45}. We leveraged the “ensemble effect” whereby gradients were averaged from three different highly accurate perceptual networks to produce a more accurate backpropagation update \cite{RN68} for the image processor network. High-level feature representations were extracted from the final convolutional layer of EfficientNet-B4, while both intermediate and high level feature representations were extracted from residual block 2, 4, 6, and 8 for WideResnet101\_2 and Resnext101\_32x8d for computation of content and style losses. Each feature representation of a processed image was compared to (using MSE) the feature representation of its corresponding multi-frame image. These comparisons were then included in a custom loss function that we describe in the next section.

\subsection{Loss Function for Training the Shadow Detector and Image Processor networks}
We successfully trained the image processor network and simultaneously removed speckle noise and retinal shadows using a combination of different loss functions. These losses were:

\subsubsection{Shadow Loss}
The shadow loss was defined to ensure that all shadows were effectively removed so that they become indistinguishable from surrounding tissues. When a given image X had been processed, it was passed to the shadow detector network to produce a predicted shadow mask, $M_\text{X}$ (with maximum pixel intensities equal to 1). All pixel intensities were then summed, and then normalized by dividing this sum with the sum of the pixels within the ground truth manually segmented mask. This normalized sum was defined as the shadow loss.

\subsubsection{Content Loss}
We used the content loss to ensure that critical information within all non-shadowed regions of a given image was retained after shadow correction. To compute content loss, we compared intermediate and high-level feature representations between a given processed image D and its corresponding multi-frame image C. Note that the content loss had been used in Style Transfer \cite{RN46} with great success at maintaining fine details and edges. We first applied the manually segmented shadow mask to the processed image, $D_\text{masked}$ and its corresponding multi-frame image, $C_\text{masked}$. This blocked out pixels in the retinal shadows so that the content loss would not be affected by any shadow removed. Next, we extracted feature representations from all perceptual networks for the processed image, and its corresponding multi-frame image. The content loss was then defined as:

\begin{equation} \label{eq:1}
L_\text{content} (D_\text{{masked}},C_\text{{masked}})= \sum\limits_{i=2,4,6,8}\frac{1}{C_{i} H_{i} W_{i}}  \lvert P_{i} (D_\text{masked} )-P_{i} (C_\text{masked} )\rvert^{2}
\end{equation}

\noindent where $P_i$ is a feature representation of the ith selected residual block of a perceptual network. Note that  i=2,4,6,8 for the WideResnet101\_2 and Resnext101\_32x8d perceptual network and i refers to the last convolutional layer of the EfficientNet-B4 perceptual network.

\subsubsection{Style Loss}
To ensure that image textures remained the same in non-shadowed regions after shadow correction, we computed the style loss for masked processed image $D_\text{masked}$ and its corresponding multi-frame image, $C_\text{masked}$. To compute the style loss, we first calculated the Gram matrix of an image to find a representation of its style. Then, the style loss for each image pair ($D_\text{masked}$, $C_\text{masked}$) was defined to be the Euclidian norm between its Gram matrices:

\begin{equation} \label{eq:2}
L_\text{style} (D_\text{masked},C_\text{masked})= \sum_{i} \lvert G_{i} (D_\text{masked} )-G_{i} (C_\text{masked} )\rvert^{2}
\end{equation}

\noindent where $G_i(x)$ is a $C_i \times C_i$ matrix defined as:
\begin{equation} \label{eq:3}
G_{i} (x)= P_{i}(x)_{C_iW_iH_i} \times P_i(x)_{H_iW_iC_i}
\end{equation}

\subsubsection{Total Loss}
The total loss was computed as a weighted sum of the content, style, and shadow losses to ensure all losses were of the same order of magnitude. The shadow loss was set as the reference (as already being normalized) and no weight was assigned. The total loss was defined as: 

\begin{equation} \label{eq:4}
L_\text{total} = \sum_{j} (w_j L_\text{content,j} + k_jL_\text{style,j}) + L_\text{shadow}
\end{equation}

\noindent where $w_j$ and $k_j$ are the weights to be derived experimentally; j summed over the type of perceptual network, i.e. EfficientNet-B4, WideResnet101\_2, and Resnext101\_32x8d. To obtain the weight values, we first trained the image processor network without style loss (k=0) to determine all w. We then introduced all style losses and normalized them so that their magnitudes were on the same scale as the content losses. Through this process the weights $w_\text{EfficientNet-B4}$, $w_\text{WideResnet101\_2}$, $w_\text{Resnext101\_32x8d}$, $k_\text{EfficientNet-B4}$, $k_\text{WideResnet101\_2}$, $k_\text{Resnext101\_32x8d}$, were given the following values: $2.86,4,6.67,6.67 \times 10^{-5}$,$1.8 \times 10^{-5}$, $2.1 \times 10^{-5}$ respectively. 

\subsection{Training Parameters}
We used 2328 multi-frame averaged B-scans during training and 300 single-frame B-scans with its corresponding multi-frame B-scan during testing. These multi-frame images were used as the ground truth images for the content and style losses, but not for the shadow loss such images still contained shadows. Each B-scan was added with a randomly generated Gaussian noise model (created according to section F). During training, the image processor network learnt how to remove the randomly generated Gaussian noise using content and style losses, and it simultaneously learnt to remove retinal blood vessel shadows through the use of the shadow loss.\\

All training and testing were performed on five Nvidia GTX 1080 Ti cards with CUDA V10.1.105, paired with Nvidia driver V436.48 and cuDNN v7.6.5. Using these hardware specifications, each image took an average of 10.3 ms to be processed. The total training time was 4 days using the Adam optimizer at a learning rate of 1 × 10–5 and a batch size of 6. A learning rate decay was implemented to halve learning rates every 10 epochs. We stopped the training when no improvements in output images could be observed.

\subsection{Noise and Retinal Shadow Removal Metrics}
We used average gradient magnitudes (AGM), the peak-signal-to-noise-ratio (PSNR), the contrast-to-noise-ratio (CNR) and the mean-structural-similarity-index (SSIM) to quantify the noise removal capabilities of our proposed algorithm. All noise removal metrics were normalized with respect to their corresponding multi-frame image for easy comparison. All noise removal metrics were extracted from regions of interest (ROIs) that did not contain retinal shadows to prevent shadow removal from affecting noise removal metrics. We also used the intra-layer contrast (ILC) and the layer-wise pixel intensity (LPI) profiles to assess the proposed algorithm’s effectiveness in removing shadows. During testing, we obtained all metrics on noisy, non-averaged single-frame B-scans. All multi-frame B-scans were then aligned to their corresponding single-frame B-scan using rigid translation/rotation transformations using 3D software (Amira, version 5.6; FEI) before noise and shadow removal metrics were extracted.

\subsubsection{Noise Removal Quantitative Assessment}
The AGM was used to quantify the sharpness of output images. We used the AGM implementation found in the Python package Numpy \cite{RN37}, defined as:

\begin{equation}\label{eq:5}
    AGM = \frac{1}{H \times W} \sum_{x}\sum_y \frac{G(x,y)}{\sqrt{2}}
\end{equation}

\noindent where $G(x,y)$,$H$ and $W$ were the gradient vector, height and width of the B-scan respectively.\\

The PSNR (expressed in dB) was used to quantify the noise levels in an image relative to its true signal strength. We used the scikit-image implementation of PSNR defined as:

\begin{equation}\label{eq:6}
    PSNR = -10 \times \log_{10} \frac{\lvert f_0 - \tilde f\rvert^2} {\lvert f_0 \rvert^2}
\end{equation}

\noindent where $f_0$ was the pixel-intensity values of the registered multi-frame B-scan, and $\tilde f$ was the pixel-intensity of the processed B-scan. A higher PSNR suggested that the processed images contained less noise and were of higher quality than images with lower PSNR.\\

The CNR provided an indication of how visible a retinal tissue layer is. It was defined as:

\begin{equation}\label{eq:7}
    CNR = \frac{\lvert \mu_r - \mu_b \rvert}{\sqrt{0.5 \times (\sigma_r^2 + \sigma_b^2)}}
\end{equation}

\noindent where $\mu_r, \mu_b, \sigma_r^2 and \sigma_b^2$ represented the means and variances of pixel intensities for a selected ROI within the tissue ‘i’ and a randomly chosen ROI from the background,  respectively. Each ROI was chosen as a $20 \times 384$ pixels region at the top of the selected B-scan. A higher CNR suggested superior visibility of tissue ‘i’ within a given B-scan. We computed the CNR for the RNFL and compared them between single-frame, processed, and multi-frame B-scans. We computed the CNR as a mean of 25 randomly selected ROIs per tissue for each given B-scan, each of size $8 \times 8$ pixels. All ROIs were manually chosen in each tissue by an expert observer (HC) using a custom Python script using the OpenCV \cite{RN75} package.\\

The SSIM was computed to quantify changes in tissue structures (i.e., edges) between a given single-frame/processed image with its corresponding multi-frame image as a reference. The SSIM was based on the computation of three terms: luminance, contrast, and structure, respectively. We used the implementation of the SSIM in the scikit-image package in Python defined as:

\begin{equation}\label{eq:8}
    SSIM(x,y) = \frac{(2\mu_x\mu_y + C_1)}{(\mu_x^2 + \mu_y^2 + C_1)(\sigma_x^2 + \sigma_y^2 + C_2)}
\end{equation}

\noindent where $\sigma_x, \mu_y, \sigma_x, \sigma_y, and \sigma_{xy}$ were the local means, standard deviations, and cross-covariance for images x, y.

\subsubsection{Shadow Removal Quantitative Assessment}
We computed the ILC to assess the performance of the proposed algorithm in removing shadows. The ILC was defined as:

\begin{equation}\label{eq:9}
    ILC = \lvert\frac{I_1 - I_2}{I_1 + I_2}\rvert
\end{equation}

\noindent where $I_1$ was the mean pixel intensity from five manually selected ROIs (size 5 ×5 pixels) that are shadow free in a given retinal layer, and $I_2$ was the corresponding value from five neighboring shadowed regions of the same tissue layer. The ILC ranged between 0 and 1, where values close to 0 indicated the absence of retinal shadows and values close to 1 indicated strongly visible blood vessel shadows.\\

We computed the intralayer contrast for multiple tissue layers of the ONH region, namely the RNFL, the photoreceptor layer (PR) and the retinal pigment epithelium (RPE) – before and after application of the proposed algorithm. Results for all metrics were recorded in the form of mean $\pm$ standard deviation.

\section{RESULTS}
\begin{figure}[H] 
    \centering
    \includegraphics[width=\textwidth,height=\textheight,keepaspectratio]{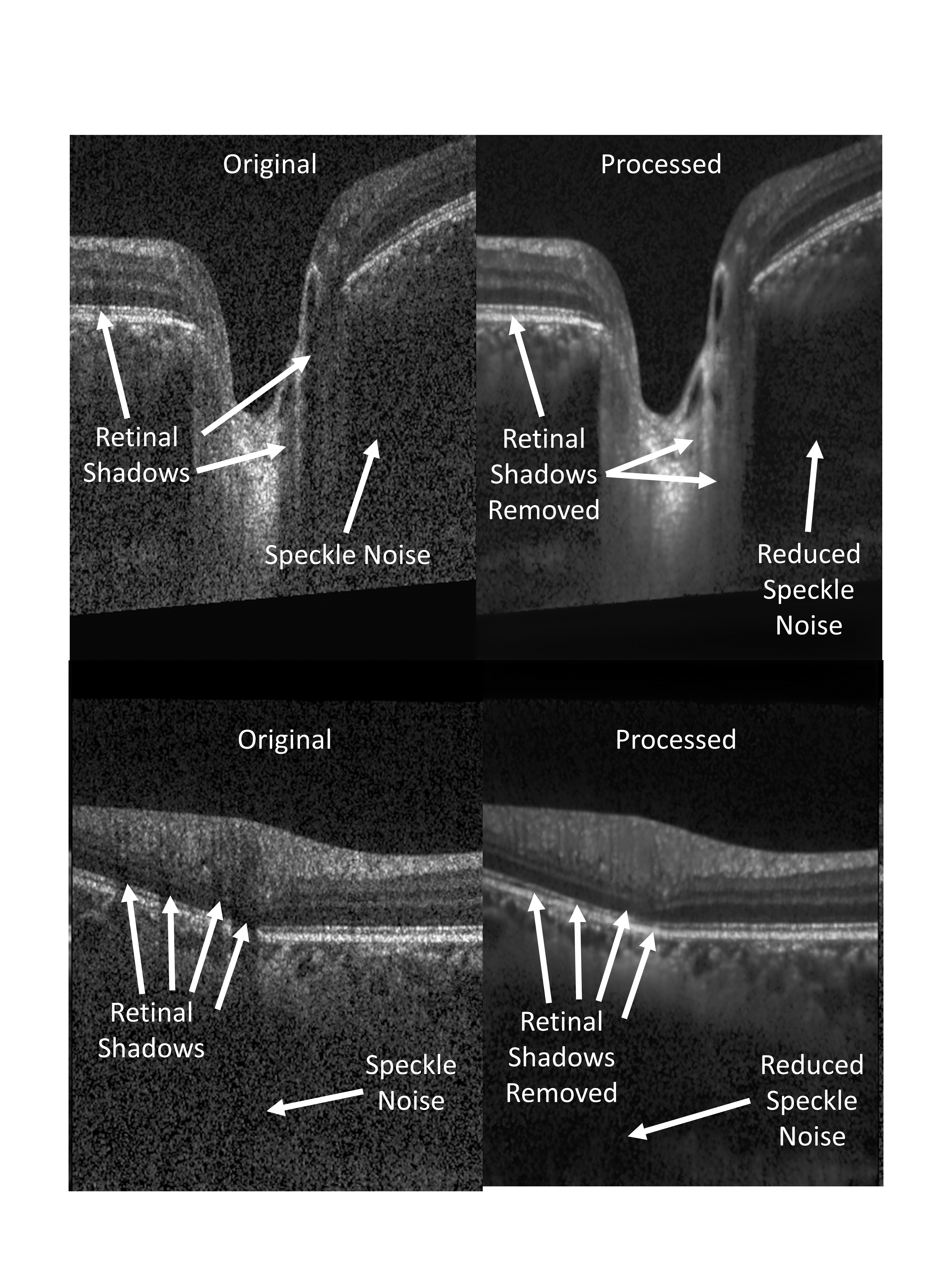}
    \caption{Samples of typical B-scans before (left) and after (right) being processed by our algorithm.}
    \label{fig:3}
\end{figure}

When trained on 2328 multi-frame B-scans with online data augmentation, our deep learning framework was able to successfully remove noise and retinal shadows from unseen single-frame B-scans (Fig. \ref{fig:3}). An independent test set of 300 single-frame B-scans was used to evaluate the noise and retinal shadow removal performance of the proposed deep learning framework qualitatively and quantitatively. The mean PSNR, the CNR and the SSIM increased with respect the input single-frame B-scans from $18.5 \pm 0.46$ dB to $20.5 \pm 0.38$ dB, $3.66 \pm 0.92$ to $8.97 \pm 2.60$ and $0.177 \pm 0.004$ to $0.45 \pm 0.09$ respectively. The ILC for the RNFL, the PR, and the RPE decreased from $0.362 \pm 0.133$ to $0.142 \pm 0.102$, $0.449 \pm 0.116$ to $0.090 \pm 0.077$, $0.381 \pm 0.100$ to $0.059 \pm 0.045$ respectively.

\subsection{Denoising Performance – Qualitative Analysis}
\begin{figure}[H] 
    \centering
    \includegraphics[width=\textwidth,height=\textheight,keepaspectratio]{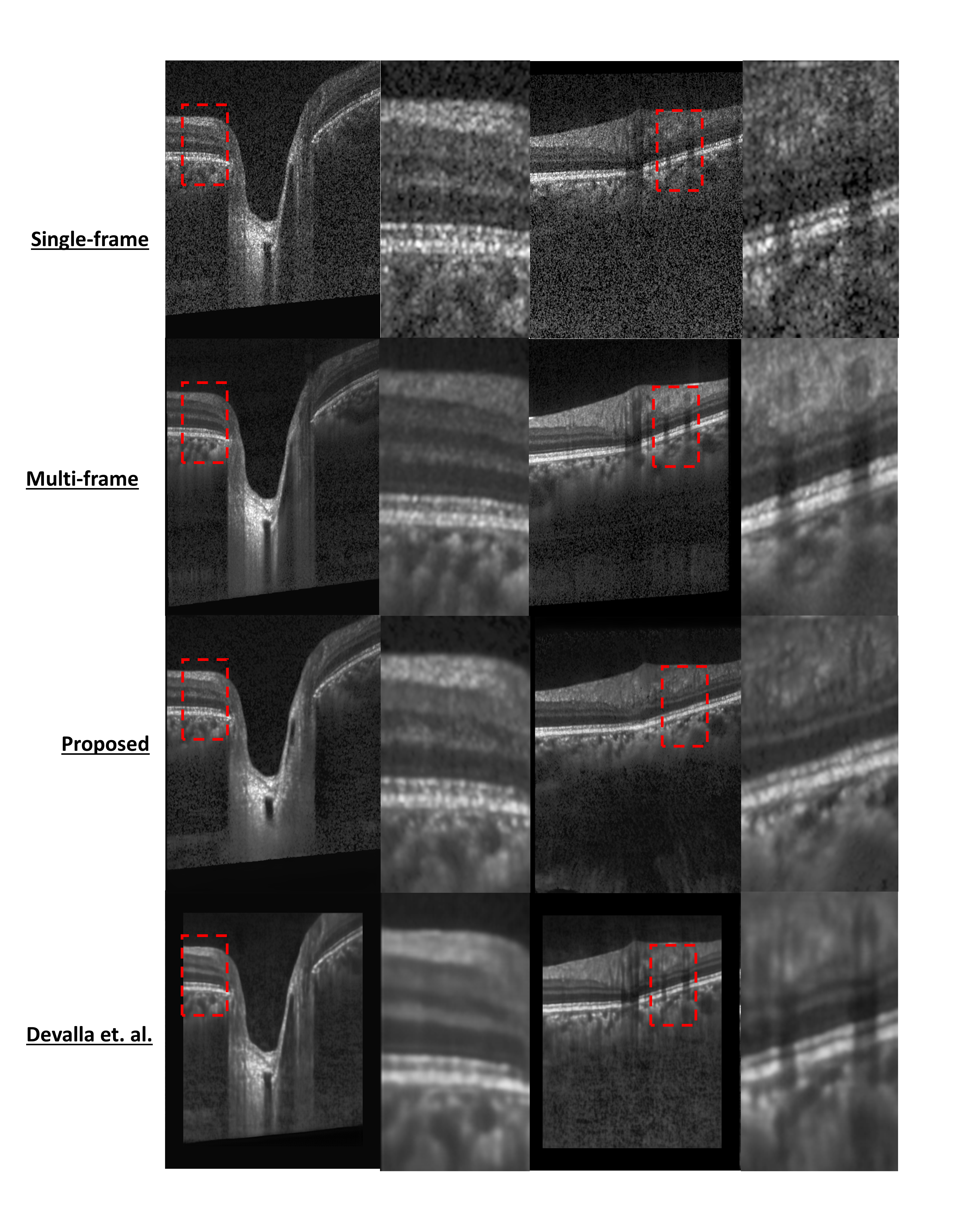}
    \caption{Qualitative analysis of the proposed noise removal and shadow removal algorithm. Blurring can be seen in the current state-of-the-art (second row) \cite{RN32}.}
    \label{fig:4}
\end{figure}

The proposed algorithm produced images without common artifacts of images processed by current deep learning frameworks, including blurring and checkerboard patterns. Single-frame B-scans processed by our algorithm looked qualitatively sharper (Fig. \ref{fig:4}) and visually closer to multi-frame B-scans than images produced by the current state-of-the-art algorithm \cite{RN32}. Retinal shadows were effectively removed, improving visible information within retinal shadows. Overall sharpness was retained and visibility of all the ONH tissues was enhanced after being processed by the proposed algorithm.

\subsection{Denoising Performance – Quantitative Analysis}
\begin{figure}[H] 
    \centering
    \includegraphics[width=\textwidth,height=\textheight,keepaspectratio]{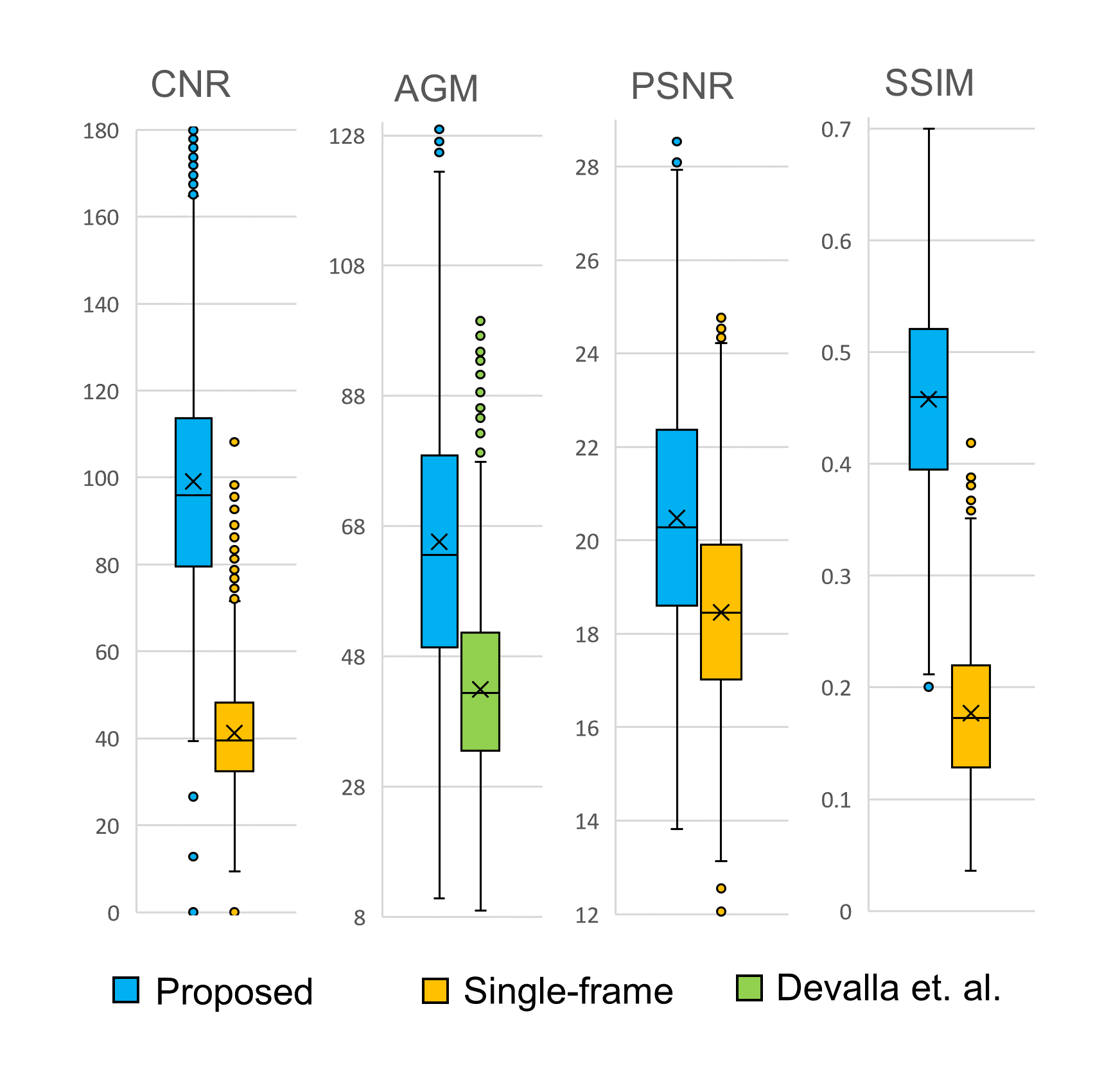}
    \caption{(from left) CNR, SSIM, and PSNR values improved relative to single-frame images. AGM of the proposed algorithm compared to that of the current state-of-the-art algorithm by Devalla et. al \cite{RN32}.}
    \label{fig:5}
\end{figure}

In Fig. \ref{fig:5}, we compared the AGM, the CNR, the PSNR, and the SSIM of images processed by the proposed algorithm with single-frame images. When evaluated on 300 single-frame B-scans, we observed that the proposed algorithm consistently produced images that are both qualitatively and quantitatively sharper than images produced by the current state-of-the-art (Fig. 4). On average, each image produced by the proposed algorithm were 154\%, 187\%, 11.1\% better than single-frame B-scans in terms of the CNR, the SSIM, the PSNR, respectively. The AGM was also 57.2\% higher than images produced by the current state-of-the-art de-noising algorithm \cite{RN32}.

\subsection{Deshadowing Performance – Quantitative Analysis}
\begin{figure}[H] 
    \centering
    \includegraphics[width=\textwidth,height=\textheight,keepaspectratio]{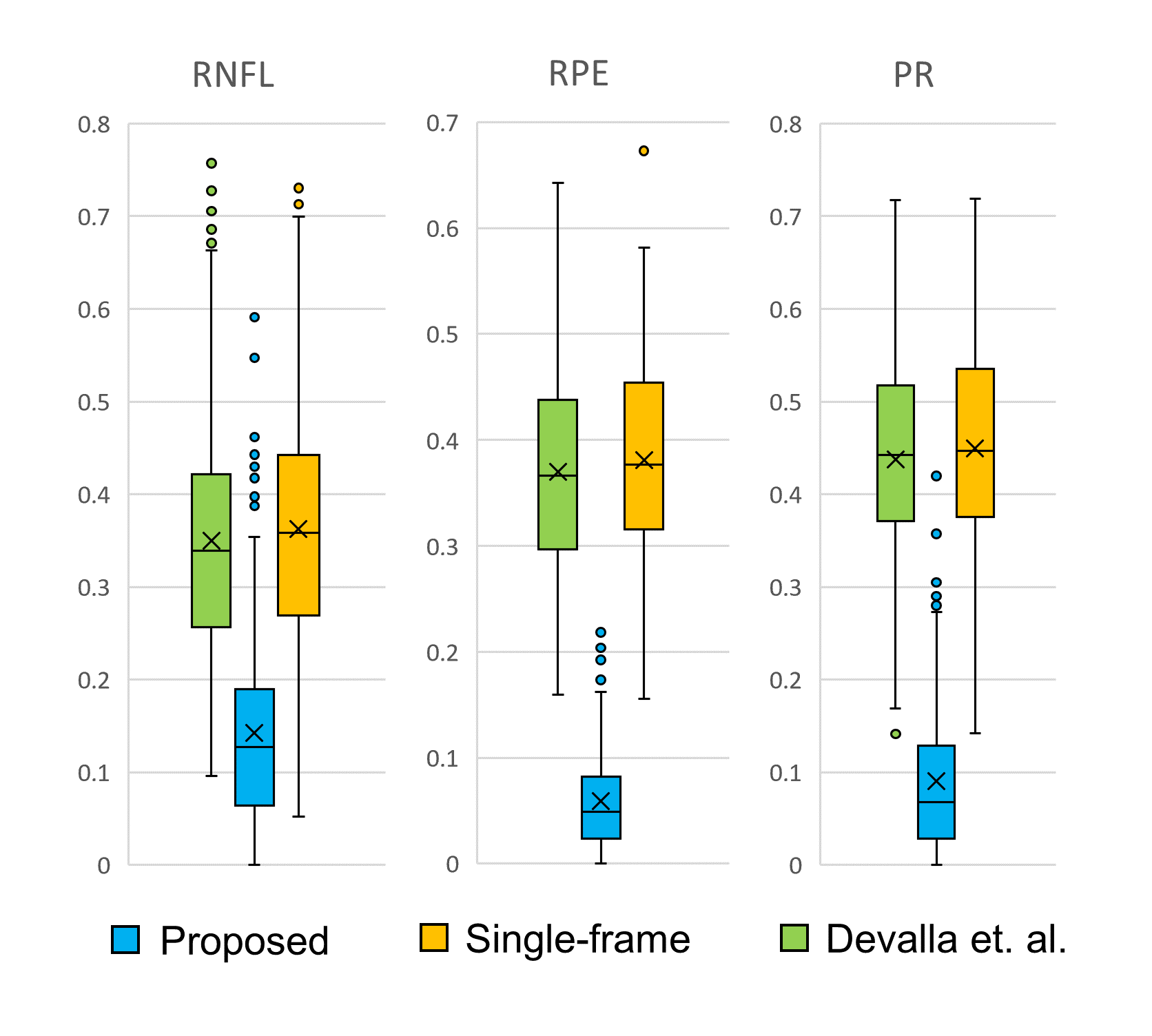}
    \caption{ILC comparison between images denoised by state-of-the-art \cite{RN32} vs proposed algorithm.}
    \label{fig:6}
\end{figure}

\begin{figure}[H] 
    \centering
    \includegraphics[width=\textwidth,height=\textheight,keepaspectratio]{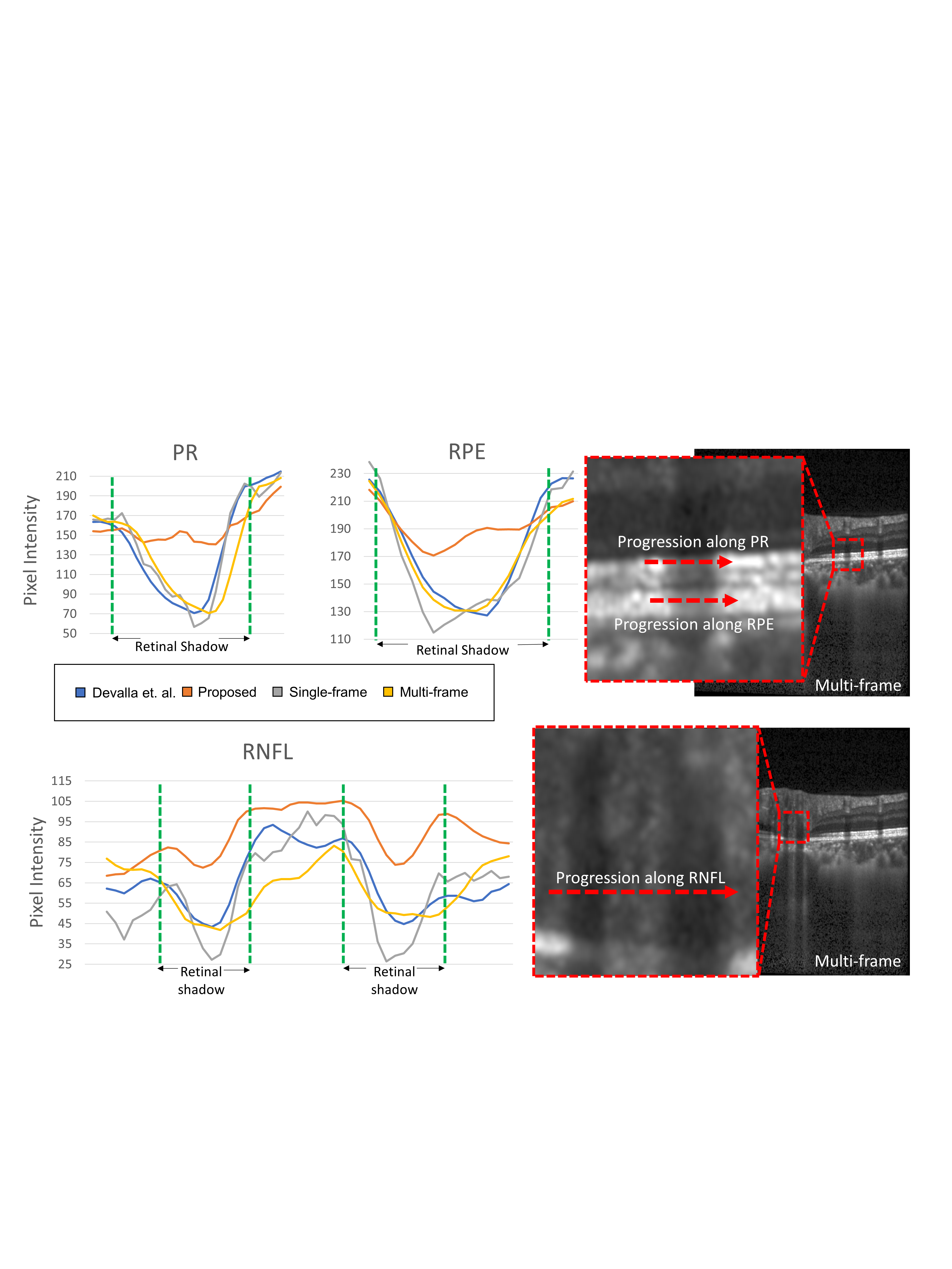}
    \caption{LPI profiles of output images along RNFL, PR and RPE layers were significantly flattener than multi-frame B-scans, or B-scans denoised by denoising state-of-the-art \cite{RN32}.}
    \label{fig:7}
\end{figure}

Our proposed algorithm produced images that had improved visibility within retinal shadows. The ILC for the RNFL, the PR, and the RPE improved by $60.0 \pm 29.3\%$, $79.0 \pm 19.4\%$ and $83.4 \pm 15.4\%$ respectively. On average, the ILC improved by $72.9 \pm 25.2\%$ (Fig. \ref{fig:6}). The LPI profiles were also significantly flattened in the RNFL, the PR and the RPE layers (Fig. \ref{fig:7}).

\section{DISCUSSION}
In this study we present a custom deep learning approach that can remove noise and retinal shadows simultaneously from single-frame OCT B-scans of the ONH. All noise removal performance metrics such as the PSNR, the CNR, and the SSIM values consistently showed significant improvements compared to single-frame images. Thus, we may be able to offer a robust deep learning framework to obtain high quality OCT B-scans with reduced scanning duration and minimized patient discomfort.\\
B-scans processed by our algorithm were qualitatively similar to their corresponding multi-frame B-scans, with the added benefit of improved visibility within retinal shadows (Fig. 3, Fig. 4). The SSIM and the CNR were significantly improved with respect to single-frame images by 154\% and 187\%, respectively. The mean AGM was also 57.2\% higher than the current state-of-the-art \cite{RN32}, providing clinicians a markedly sharper image. Image sharpness is critical given that many pathologies require sharp layer boundaries for accurate retinal layer thickness measurements. One example would be quantifying macular edema, which require measurements of retinal thickness in response to therapy \cite{RN29}. Given the significance of retinal layers and connective tissues in the prognosis and diagnosis of ocular pathologies such as glaucoma and age-related macular degeneration, enhanced visibility would improve automated algorithms downstream of the post processing pipeline, namely alignment, registration, segmentation, diagnosis and, ultimately, prognosis.\\
The proposed algorithm did not require any further segmentation, delineation, or identification of shadows by the user. Similar to our previous work \cite{RN27}, the ILC mean and standard deviation decreased with the depth of the retinal layer of interest (Fig. 6), suggesting that performance of the proposed algorithm was consistently better in deeper layers. The proposed algorithm substantially recovered the visibility of the anterior lamina cribrosa (LC) boundary and anterior LC insertion, which may result in a more confident prediction of early glaucoma \cite{RN30}. Moreover, the main load bearing tissues of the eye in the ONH region, such as the LC and adjacent peripapillary sclera, could be monitored for pre-disease biomechanical and morphological changes. Changes in these tissues have been previously identified as risk factors for glaucoma \cite{RN31}. Measurements of the anatomy of such tissues could be more robust and substantially improved after application of the proposed algorithm.\\

In this study, several limitations warrant further discussion. While we did not find any evidence of pathology being obscured or introduced into output images, it is extremely important to validate this in pathological cases. However, we would need to image the exact same tissue region with and without the presence of blood flow (to remove retinal blood vessel shadows). Such experiments would be extremely complex to carry out in vivo, especially in humans, even if blood vessels were to be flushed with saline during experiments. Such validations may be required for full clinical acceptance of this methodology. Furthermore, it would be critical to also confirm that the proposed algorithm would not interfere with another AI algorithm (especially those aimed at diagnosis and prognosis). Nevertheless, it is possible that the proposed algorithm might improve diagnosis and prognosis algorithms by improving the quality of the input data. We aim to test this hypothesis in the future.\\

Furthermore, although the proposed algorithm functioned well on single-frame images from healthy individuals, more work is required to ensure that it can reproduce similar performance on B-scans of eyes with pathophysiological conditions such as glaucoma. This is especially critical for deep learning approaches, which respond unpredictably to input data that is different from images used during training. As this algorithm was trained on single-frame images from a Spectralis OCT device, it is unknown if it can maintain this performance on OCT images from other devices. Each scenario stated above may require a separate training set. Our future studies will therefore focus on validating the performance of the proposed algorithm across devices and between healthy and pathological eyes.

\section{CONCLUSION}
The proposed algorithm successfully removed both noise and retinal shadows from single-frame B-scans. The algorithm also drastically reduced the time needed (3.5 minutes to 10.6s) for medical professionals and patients to obtain high quality post-processed B-scans. This could have significant economic benefits for hospitals by allowing less money reducing expenditure on expensive, high quality OCT machines. Patients would also benefit by a reduction in the time needed to remain in a fixated position during OCT image acquisition. Automated segmentation and diagnosis algorithms could also benefit clinical diagnostics by providing increased structural clarity, improved layer continuity and enhanced visibility both within shadows and retinal layers. The combination of both noise removal and retinal shadow removal algorithms in a single step will improve latency and be a step toward the goal of real-time OCT image processing.

\section*{FUNDING}
Supported by Singapore Ministry of Education Academic Research Funds Tier 1 (R-155-000-168-112 to AT; R-397-000-294-114 to MJAG); National University of Singapore Young Investigator Award Grants (NUSYIA FY16 P16, R-155-000-180-133 to AT; NUSYIA FY13 P03, R-397-000-174-133 to MJAG); Singapore Ministry of Education Academic Research Funds Tier 2 (R-397-000-280-112, R-397-000-308-112 to MJAG); and National Medical Research Council Grant NMRC/STAR/0023/2014 (TA).

\section*{DISCLOSURES}
Haris Cheong: None, Sripad Krishna Devalla: None, Thanadet Chuangsuwanich: None, Tin A. Tun: None, Xiaofei Wang: None, Tin Aung: None, Leopold Schmetterer: None, Martin L. Buist: None, Craig Boote: None, Alexandre H. Thiéry: Abyss Processing (Co-Founder), Michaël J. A. Girard: Abyss Processing (Co-Founder)

\bibliographystyle{unsrt}
\bibliography{denoising}

\begin{thebibliography}{10}

\bibitem{RN54}
Carmen~A Puliafito, Michael~R Hee, Charles~P Lin, Elias Reichel, Joel~S
  Schuman, Jay~S Duker, Joseph~A Izatt, Eric~A Swanson, and James~G Fujimoto.
\newblock Imaging of macular diseases with optical coherence tomography.
\newblock {\em Ophthalmology}, 102(2):217--229, 1995.

\bibitem{RN13}
Alexander Wong, Akshaya Mishra, Kostadinka Bizheva, and David~A Clausi.
\newblock General bayesian estimation for speckle noise reduction in optical
  coherence tomography retinal imagery.
\newblock {\em Optics express}, 18(8):8338--8352, 2010.

\bibitem{RN48}
Maciej Szkulmowski, Iwona Gorczynska, Daniel Szlag, Marcin Sylwestrzak, Andrzej
  Kowalczyk, and Maciej Wojtkowski.
\newblock Efficient reduction of speckle noise in optical coherence tomography.
\newblock {\em Optics express}, 20(2):1337--1359, 2012.

\bibitem{RN33}
Mitsuro Sugita, Stefan Zotter, Michael Pircher, Tomoyuki Makihira, Kenichi
  Saito, Nobuhiro Tomatsu, Makoto Sato, Philipp Roberts, Ursula
  Schmidt-Erfurth, and Christoph~K Hitzenberger.
\newblock Motion artifact and speckle noise reduction in polarization sensitive
  optical coherence tomography by retinal tracking.
\newblock {\em Biomedical optics express}, 5(1):106--122, 2014.

\bibitem{RN34}
Jing Wu, Bianca~S Gerendas, Sebastian~M Waldstein, Georg Langs, Christian
  Simader, and Ursula Schmidt-Erfurth.
\newblock Stable registration of pathological 3d-oct scans using retinal
  vessels.
\newblock In {\em Proceedings of the Ophthalmic Medical Image Analysis First
  International Workshop}, 2014.

\bibitem{RN35}
Iwona Gorczynska, Justin~V Migacz, Robert~J Zawadzki, Arlie~G Capps, and John~S
  Werner.
\newblock Comparison of amplitude-decorrelation, speckle-variance and
  phase-variance oct angiography methods for imaging the human retina and
  choroid.
\newblock {\em Biomedical optics express}, 7(3):911--942, 2016.

\bibitem{RN39}
Fei Shi, Ning Cai, Yunbo Gu, Dianlin Hu, Yuhui Ma, Yang Chen, and Xinjian Chen.
\newblock Despecnet: a cnn-based method for speckle reduction in retinal
  optical coherence tomography images.
\newblock {\em Physics in Medicine \& Biology}, 64(17):175010, 2019.

\bibitem{RN1}
Vivek~J Srinivasan, Maciej Wojtkowski, Andre~J Witkin, Jay~S Duker, Tony~H Ko,
  Mariana Carvalho, Joel~S Schuman, Andrzej Kowalczyk, and James~G Fujimoto.
\newblock High-definition and 3-dimensional imaging of macular pathologies with
  high-speed ultrahigh-resolution optical coherence tomography.
\newblock {\em Ophthalmology}, 113(11):2054--2065. e3, 2006.

\bibitem{RN58}
Atsushi Sakamoto, Masanori Hangai, and Nagahisa Yoshimura.
\newblock Spectral-domain optical coherence tomography with multiple b-scan
  averaging for enhanced imaging of retinal diseases.
\newblock {\em Ophthalmology}, 115(6):1071--1078. e7, 2008.

\bibitem{RN17}
Giovanni~J Ughi, Matilda Larsson, Christophe Dubois, Peter~R Sinnaeve, Walter
  Desmet, Jan D'Hooge, Tom Adriaenssens, and Mark Coosemans.
\newblock Automatic three-dimensional registration of intravascular optical
  coherence tomography images.
\newblock {\em Journal of biomedical optics}, 17(2):026005, 2012.

\bibitem{RN18}
Shaozhen Song, Zhihong Huang, and Ruikang~K Wang.
\newblock Tracking mechanical wave propagation within tissue using
  phase-sensitive optical coherence tomography: motion artifact and its
  compensation.
\newblock {\em Journal of biomedical optics}, 18(12):121505, 2013.

\bibitem{RN50}
SH~Yun, GJ~Tearney, JF~De~Boer, and BE~Bouma.
\newblock Motion artifacts in optical coherence tomography with
  frequency-domain ranging.
\newblock {\em Optics Express}, 12(13):2977--2998, 2004.

\bibitem{RN60}
Yali Jia, Steven~T Bailey, David~J Wilson, Ou~Tan, Michael~L Klein, Christina~J
  Flaxel, Benjamin Potsaid, Jonathan~J Liu, Chen~D Lu, and Martin~F Kraus.
\newblock Quantitative optical coherence tomography angiography of choroidal
  neovascularization in age-related macular degeneration.
\newblock {\em Ophthalmology}, 121(7):1435--1444, 2014.

\bibitem{RN23}
Cong Ye, Marco Yu, and Christopher~Kai‐shun Leung.
\newblock Impact of segmentation errors and retinal blood vessels on retinal
  nerve fibre layer measurements using spectral‐domain optical coherence
  tomography.
\newblock {\em Acta Ophthalmologica}, 94(3):e211--e219, 2016.

\bibitem{RN24}
Jehn-Yu Huang, Melike Pekmezci, Nisreen Mesiwala, Andrew Kao, and Shan Lin.
\newblock Diagnostic power of optic disc morphology, peripapillary retinal
  nerve fiber layer thickness, and macular inner retinal layer thickness in
  glaucoma diagnosis with fourier-domain optical coherence tomography.
\newblock {\em Journal of glaucoma}, 20(2):87--94, 2011.

\bibitem{RN2}
Michaël~JA Girard, Nicholas~G Strouthidis, C~Ross Ethier, and Jean~Martial
  Mari.
\newblock Shadow removal and contrast enhancement in optical coherence
  tomography images of the human optic nerve head.
\newblock {\em Investigative ophthalmology \& visual science},
  52(10):7738--7748, 2011.

\bibitem{RN20}
Xiaojiao Mao, Chunhua Shen, and Yu-Bin Yang.
\newblock Image restoration using very deep convolutional encoder-decoder
  networks with symmetric skip connections.
\newblock In {\em Advances in neural information processing systems}, pages
  2802--2810, 2016.

\bibitem{RN21}
Yuhui Ma, Xinjian Chen, Weifang Zhu, Xuena Cheng, Dehui Xiang, and Fei Shi.
\newblock Speckle noise reduction in optical coherence tomography images based
  on edge-sensitive cgan.
\newblock {\em Biomedical optics express}, 9(11):5129--5146, 2018.

\bibitem{RN32}
Sripad~Krishna Devalla, Giridhar Subramanian, Tan~Hung Pham, Xiaofei Wang,
  Shamira Perera, Tin~A. Tun, Tin Aung, Leopold Schmetterer, Alexandre~H.
  Thiéry, and Michaël J.~A. Girard.
\newblock A deep learning approach to denoise optical coherence tomography
  images of the optic nerve head.
\newblock {\em Scientific Reports}, 9(1):14454, 2019.

\bibitem{RN40}
Neha Gour and Pritee Khanna.
\newblock Speckle denoising in optical coherence tomography images using
  residual deep convolutional neural network.
\newblock {\em Multimedia Tools and Applications}, pages 1--17, 2019.

\bibitem{RN41}
Min Xu, Chen Tang, Fugui Hao, Mingming Chen, and Zhenkun Lei.
\newblock Texture preservation and speckle reduction in poor optical coherence
  tomography using the convolutional neural network.
\newblock {\em Medical Image Analysis}, page 101727, 2020.

\bibitem{RN42}
Zailiang Chen, Ziyang Zeng, Hailan Shen, Xianxian Zheng, Peishan Dai, and
  Pingbo Ouyang.
\newblock Dn-gan: Denoising generative adversarial networks for speckle noise
  reduction in optical coherence tomography images.
\newblock {\em Biomedical Signal Processing and Control}, 55:101632, 2020.

\bibitem{RN26}
Kiran~Kumar Vupparaboina, Kunal~K Dansingani, Abhilash Goud, Mohammed~Abdul
  Rasheed, Fayez Jawed, Soumya Jana, Ashutosh Richhariya, K~Bailey Freund, and
  Jay Chhablani.
\newblock Quantitative shadow compensated optical coherence tomography of
  choroidal vasculature.
\newblock {\em Scientific reports}, 8(1):1--9, 2018.

\bibitem{RN27}
Haris Cheong, Sripad~Krishna Devalla, Tan~Hung Pham, Liang Zhang, Tin~Aung Tun,
  Xiaofei Wang, Shamira Perera, Leopold Schmetterer, Tin Aung, and Craig Boote.
\newblock Deshadowgan: a deep learning approach to remove shadows from optical
  coherence tomography images.
\newblock {\em Translational Vision Science \& Technology}, 9(2):23--23, 2020.

\bibitem{RN63}
Qinqin Zhang, Fang Zheng, Elie~H Motulsky, Giovanni Gregori, Zhongdi Chu,
  Chieh-Li Chen, Chunxia Li, Luis De~Sisternes, Mary Durbin, and Philip~J
  Rosenfeld.
\newblock A novel strategy for quantifying choriocapillaris flow voids using
  swept-source oct angiography.
\newblock {\em Investigative ophthalmology \& visual science}, 59(1):203--211,
  2018.

\bibitem{RN69}
Ian Wong, Hideki Koizumi, and Wico Lai.
\newblock Enhanced depth imaging optical coherence tomography.
\newblock {\em Ophthalmic surgery, lasers \& imaging : the official journal of
  the International Society for Imaging in the Eye}, 42 Suppl:S75--84, 2011.

\bibitem{RN70}
R~Daniel Ferguson, Daniel~X Hammer, Lelia~Adelina Paunescu, Siobahn Beaton, and
  Joel~S Schuman.
\newblock Tracking optical coherence tomography.
\newblock {\em Optics letters}, 29(18):2139--2141, 2004.

\bibitem{RN72}
Daniel~X Hammer, R~Daniel Ferguson, Nicusor~V Iftimia, Teoman Ustun, Gadi
  Wollstein, Hiroshi Ishikawa, Michelle~L Gabriele, William~D Dilworth, Larry
  Kagemann, and Joel~S Schuman.
\newblock Advanced scanning methods with tracking optical coherence tomography.
\newblock {\em Optics express}, 13(20):7937--7947, 2005.

\bibitem{RN64}
Olaf Ronneberger, Philipp Fischer, and Thomas Brox.
\newblock U-net: Convolutional networks for biomedical image segmentation.
\newblock In {\em International Conference on Medical image computing and
  computer-assisted intervention}, pages 234--241. Springer, 2015.

\bibitem{RN65}
Shie Mannor, Dori Peleg, and Reuven Rubinstein.
\newblock The cross entropy method for classification.
\newblock In {\em Proceedings of the 22nd international conference on Machine
  learning}, pages 561--568, 2005.

\bibitem{RN66}
Matthew~D Zeiler, M~Ranzato, Rajat Monga, Min Mao, Kun Yang, Quoc~Viet Le,
  Patrick Nguyen, Alan Senior, Vincent Vanhoucke, and Jeffrey Dean.
\newblock On rectified linear units for speech processing.
\newblock In {\em 2013 IEEE International Conference on Acoustics, Speech and
  Signal Processing}, pages 3517--3521. IEEE, 2013.

\bibitem{RN43}
Mingxing Tan and Quoc~V Le.
\newblock Efficientnet: Rethinking model scaling for convolutional neural
  networks.
\newblock {\em arXiv preprint arXiv:1905.11946}, 2019.

\bibitem{RN44}
Sergey Zagoruyko and Nikos Komodakis.
\newblock Wide residual networks.
\newblock {\em arXiv preprint arXiv:1605.07146}, 2016.

\bibitem{RN45}
Saining Xie, Ross Girshick, Piotr Dollár, Zhuowen Tu, and Kaiming He.
\newblock Aggregated residual transformations for deep neural networks.
\newblock In {\em Proceedings of the IEEE conference on computer vision and
  pattern recognition}, pages 1492--1500, 2016.

\bibitem{RN68}
Sean Tao.
\newblock Deep neural network ensembles.
\newblock In {\em International Conference on Machine Learning, Optimization,
  and Data Science}, pages 1--12. Springer, 2019.

\bibitem{RN46}
Leon~A Gatys, Alexander~S Ecker, and Matthias Bethge.
\newblock Image style transfer using convolutional neural networks.
\newblock In {\em Proceedings of the IEEE conference on computer vision and
  pattern recognition}, pages 2414--2423, 2016.

\bibitem{RN37}
Travis~E Oliphant.
\newblock {\em A guide to NumPy}, volume~1.
\newblock Trelgol Publishing USA, 2006.

\bibitem{RN75}
Gary Bradski and Adrian Kaehler.
\newblock Opencv.
\newblock {\em Dr. Dobb’s journal of software tools}, 3, 2000.

\bibitem{RN29}
Glenn~J Jaffe and Joseph Caprioli.
\newblock Optical coherence tomography to detect and manage retinal disease and
  glaucoma.
\newblock {\em American journal of ophthalmology}, 137(1):156--169, 2004.

\bibitem{RN30}
Kyoung~Min Lee, Tae-Woo Kim, Robert~N Weinreb, Eun~Ji Lee, Michaël~JA Girard,
  and Jean~Martial Mari.
\newblock Anterior lamina cribrosa insertion in primary open-angle glaucoma
  patients and healthy subjects.
\newblock {\em PLoS One}, 9(12):e114935, 2014.

\bibitem{RN31}
Hongli Yang, J~Crawford Downs, Christopher Girkin, Lisandro Sakata, Anthony
  Bellezza, Hilary Thompson, and Claude~F Burgoyne.
\newblock 3-d histomorphometry of the normal and early glaucomatous monkey
  optic nerve head: lamina cribrosa and peripapillary scleral position and
  thickness.
\newblock {\em Investigative ophthalmology \& visual science},
  48(10):4597--4607, 2007.

\end{thebibliography}

\end{document}